\documentclass{mn2e}
\usepackage{psfig}
\newcommand{\beq}{\begin{equation}}
\newcommand{\eeq}{\end{equation}}
\newcommand{\bea}{\begin{eqnarray}}
\newcommand{\eea}{\end{eqnarray}}
\def\ltsima{$\; \buildrel < \over \sim \;$}
\def\lsim{\lower.5ex\hbox{\ltsima}}
\def\gtsima{$\; \buildrel > \over \sim \;$}
\def\gsim{\lower.5ex\hbox{\gtsima}}
\begin{document}
\title{Jitter Radiation In Gamma Ray Bursts and their afterglows:
Emission and Self Absorption} 
\author[Workman et al.]{Jared C. Workman$^1$, Brian J. Morsony$^1$,
Davide Lazzati$^1$ and Mikhail V. Medvedev$^2$ \\
$^1$ JILA, University of Colorado, 440 UCB, Boulder, CO 80309-0440, USA \\
$^2$ Department of Physics and Astronomy, University of Kansas, 
Lawrence, KS 66045}

\maketitle

\begin{abstract}
Relativistic electrons moving into a highly tangled magnetic field
emit jitter radiation. We present a detailed computation of the jitter
radiation spectrum, including self-absorption, for electrons inside
Weibel-like shock generated magnetic fields. We apply our results to
the case of the prompt and afterglow emission of gamma-ray bursts. We
show that jitter emission can reproduce most of the observed features
with some important differences with respect to standard synchrotron,
especially in the frequency range between the self-absorption and the
peak frequency. We discuss the similarities and differences between
jitter and synchrotron and discuss experiments that can disentangle
the two mechanisms.
\end{abstract}


\section{Introduction}

The external shock model (Meszaros \& Rees 1997; Piran 1999) has been
very successful in the explanation of afterglow radiation in gamma-ray
bursts. In this model the afterglow photons are produced by converting
some of the internal energy of the burst blastwave into radiation. The
radiation mechanism is supposed to be synchrotron produced by a
population of relativistic electrons gyrating in a high intensity
magnetic field. Synchrotron radiation is able to reproduce the spectra
of some observed GRB afterglows (Wijers et al. 1999; Panaitescu \&
Kumar 2001; Panaitescu 2005), their temporal decays, their duration
and their small level of linear polarization (Covino et al. 1999;
Lazzati et al. 2004).

How the magnetic field is produced is, however, a matter of open
debate. Compression of the interstellar magnetic field would generate
a field that is not large enough to explain the observed
frequencies. On the other hand, a strong magnetic field linked to the
central object powering the explosion (e.g. the magnetic field of a
neutron star) decays too rapidly with radius. In the standard external
shock model, it is assumed that a quasi-equipartition magnetic field
is generated by the shock and permeates the shocked interstellar
medium (ISM) region without decaying. The fraction of energy given to
the magnetic field is usually defined through
$B^2=8\pi\epsilon_B\,\rho$ where $\rho$ is the energy density of the
post-shock material and $\epsilon_B$ is a non-dimensional parameter
called the magnetic field equipartition parameter. Synchrotron
radiation is also supposed to generate the prompt emission photons. The
synchrotron interpretation of prompt GRB spectra has however been
unable to explain a sizable fraction of the spectra that display
low-energy slopes steeper than $\nu^{1/3}$, i.e., harder than what is
allowed by synchrotron. In addition, models in which the electrons are
accelerated impulsively suffer from the cooling problem, i.e., while
the cooling time of the electrons is extremely short, no sign of an
aging electron population is observed in the spectra (Imamura \&
Epstein 1987; Crider et al. 1997; Preece et al. 1998; Ghisellini et
al. 2000). In this paper we do not attempt to solve the cooling
problem, and implicitly assume that it can be solved by allowing for a
slow or repeated acceleration of the emitting electrons.

Particle in cell (PIC) simulations of collisionless shocks have
demonstrated that the Weibel instability can produce magnetic fields
at almost equipartition ($\epsilon_B\sim0.1$, Silva et al. 2003;
Fredriksen et al. 2004). Recent state-of-the-art PIC simulations
demonstrate that the Weibel-generated magnetic fields survive for at
least a few hundred skin depths (Spitkovsky 2005; Chang, et al. 2007;
Spitkovsky 2007), in agreement with predictions by Medvedev, et
al. (2005). Whether these fields can avoid dissipation and survive
beyond the few hundred plasma skin depths remains an open question.
What appears certain is that the field created by the Weibel
instability has a very short coherence length, so that classical
synchrotron formulae cannot be generally applied.

Radiation produced by relativistic particles in small-scale magnetic
fields has been a subject of study for several decades. Historically,
it begins with two seminal papers by Landau \& Pomeranchuk (1953) and
Migdal (1954), who predicted the suppression of the spectral power of
low-energy harmonics --- the Landau-Pomeranchuk-Migdal
effect. Interestingly, the spectral properties of synchrotron
radiation from either homogeneous fields or from random fields
coherent on scales larger than the particle Larmor radius is rather
universal, as it can be straightforwardly applied in a wide variety of
magnetic field configurations. In contrast to synchrotron radiation,
the radiation emitted from a magnetic field (not necessarily random)
whose gradient scale is smaller than the particle gyro-orbit, is
determined by the particular field configuration.  Therefore, one has
to be very specific about the field structure in the system for which
the radiation spectrum is computed.

A number of theories of radiation by relativistic electrons in
small-scale magnetic fields in various regimes have been developed,
including radiation in wiggler/undulator fields and free-electron
lasers (presently used in sources of X-ray and extreme UV light)
(Kincaid 1977; Joshi et al. 1987; Fedele et al. 1990; Williams et
al. 1993; Attwood et al. 1993, Attwood 2000), radiation emitted by
cosmic rays in the interstellar medium plasma turbulence if it extends
to sufficiently small scales, beyond the Larmor radius of thermal
electrons (Kaplan \& Tsytovich 1969; Bel'kov et al 1980; Ginzburg \&
Tsytovich 1984; Toptygin 1985; Toptygin \& Fleishman 1987ab);
transition radiation in inhomogeneous plasmas (for a good review, see
monograph by Ginzburg \& Tsytovich 1984); synchro-Compton (or
nonlinear inverse Compton) radiation produced by electrons in the
non-stationary field of a strong electromagnetic wave (Gunn \&
Ostriker 1971; Rees 1971ab; Blandford 1972); betatron radiation
emitted by a particle propagating in a magnetic field of an ion
current channel (somewhat analogous to a Weibel current filament;
Wang, et al 2002).

Medvedev (2000) considered radiation from Weibel-generated
fields, generalizing the wiggler/undulator radiation to the case of
randomly distributed current filaments and associated fields as a
model of a GRB shock. The spectral properties of radiation have been
analytically derived, using the perturbation theory (Landau \&
Lifschitz 1971), for a single electron and for a power-law electron
energy distribution, in the limit of a magnetic field with a very
small coherence length.  The theory of jitter radiation applied to the
prompt emission of GRBs was shown to be able to reproduce some
characteristic features, such as the steep low-energy spectrum and the
sharpness of the spectral break (Medvedev 2000). The theory has been
further generalized to the 3D case by Fleishman (2006b) who calls it
``Diffusive Synchrotron Radiation'' (DSR) and by Medvedev
(2006). Jitter radiation and diffusive synchrotron radiation describe
the same phenomenon: radiation from particles traveling through small
scale, random magnetic fields coherent on the scale of a plasma skin
depth.  The primary difference in the two approaches is that Medvedev
uses the perturbative approach introduced by Landau \& Lifschitz (1971)
whereas Fleishman applied both the perturbative treatment and a
more general non-perturbative scheme which allows for the effects of
scattering to appear in the spectra as the coherence length of the
field increase.  Fleishman (2006b) and, independently, Medvedev
(2006) obtained the single electron spectrum to be $F_\nu\propto\nu^0$
in the perturbative approach, noting the possibility that steeper
spectra (up to $\nu\propto\nu^1$) could be obtained if the magnetic
field spectrum could be factorized. Previous results were obtained by
solving the non-perturbative equation describing single electrons
emitting radiation due to both large scale ordered fields and small
scale random fields (Toptygin 1985; Toptygin \& Fleishman 1987a).  The
anisotropy of jitter radiation has also been suggested as a possible
explanation of the rapid spectral variability of prompt GRB emission
(Medvedev 2006). Recently, the jitter radiation model has been
extended to the self-absorption regime (Medvedev, et al. 2007) and
applied to GRB afterglows. Approximated analytical spectra and light
curves were computed.

In this paper we use the theory of jitter radiation as it is
more amenable to numerical and analytical calculations (following
Medvedev 2006) to improve upon the analytical computations of Medvedev
et al. 2007. We review and refine the jitter theory including
self-absorption, numerically implement it in a radiation code, and
apply it for the optical-UV range for the prompt emission of GRBs. We
also compare properties and spectral characteristics of jitter and
synchrotron radiation models in the afterglow regime. We show that
jitter radiation is a viable framework for the interpretation of
prompt and afterglow observations. We discuss the observational
implications of this alternative radiation mechanism and discuss some
observational tests to compare synchrotron and jitter radiation in
GRBs.

The paper is organized as follows: in \S~2 we compute the emissivity
of jitter radiation and the absorption coefficient, in \S~3 we discuss
the range over which our results are valid.  In \S~4 we apply our
result to some cases relevant for GRB prompt and afterglow
emission. We conclude in \S~5 by discussing some implications for GRB
observations.

\section{Spectrum of Jitter Radiation}

This section lays out the analytical framework we use to compute the
specific intensity $I_{\omega}$ of jitter radiation produced by Weibel
generated magnetic fields. In the first part we detail how we
mathematically describe the magnetic field produced by the Weibel
instability and used in the calculations. In the second part we
discuss the mathematical formulation used to calculate the emission
coefficient $j_{\omega}$, which describes the angle-averaged power per
unit frequency added to the radiation field by electrons emitting
jitter radiation.  Finally, the third part describes how to obtain the
absorption coefficient $\alpha_{\omega}$, which describes how jitter
radiation self attenuates.
 
As a first step we derive the absorption and emission coefficients in
the comoving frame.  Although we calculate the comoving quantities, it
is straightforward to transform to the observer frame by recalling
that $\alpha_{\omega}\omega$ and $\frac{j_{\omega}}{\omega^2}$ are
both Lorentz invariants with
$\omega=\omega'\gamma_b(1+\frac{v}{c}\cos\Theta')$ where $\gamma_{b}$
and $v$ refer to the bulk velocity of the object and not to the
Lorentz factors of the individual emitting electrons and $\cos\Theta'$
is defined by the angle between the photon path and the co-moving
volume velocity direction.  This algebraic transformation allows us to
recast the equation for the observed intensity as a function of
comoving quantities.  Proceeding in this manner greatly simplifies the
analytics and allows us to solve for the observed specific intensity
by using numerically computed, comoving quantities.

\subsection{Magnetic Fields In Gamma Ray Bursts}

The standard model for emission from both the prompt burst in GRBs and
the afterglow relies mainly on the assumption that the radiative
mechanism is synchrotron (Zhang \& Meszaros 2004; Granot et
al. 1999ab).  While the assumption of synchrotron is reasonable for
some of the observed prompt bursts due to strong magnetic field
progenitors, it is unlikely that it is always dominant in afterglow
emission.  Parameterizing the strength of the magnetic field by
$\epsilon_B$, which is the ratio of the magnetic field energy density
to the thermal energy density, leads to values of $\epsilon_B$ (from
observations and simulations) ranging from $10^{-1}$ to $10^{-5}$
(Medvedev \& Loeb 1999; Panaitescu \& Kumar 2001; Panaitescu 2005).

The field may be directly connected to the compact object powering the
GRB outflow. In that case, assuming a strong field of $\sim10^{16}$~G
on the surface of the compact object and considering volume expansion
in an outflow ($B \propto V^{-2/3}$), one obtains $B\sim 10^{-2}$~G
($\epsilon_B\sim10^{-7}$) in the afterglow region ($R\sim10^{16}$~cm).
Such a field would be too small to reproduce the observations and
would decay too steeply with time. Alternatively, the magnetic field
could be generated through shock compression of a pre-existing
interstellar field. In that case, $B\prime\propto\gamma B_{ISM}$ is
too weak to generate sufficiently strong fields at the afterglow shock
front and results in values of $\epsilon_B\sim 10^{-11}$ (Medvedev \&
Loeb 1999).  With no new mechanism available to produce fields it is
unlikely that synchrotron can always be the mechanism by which
afterglow radiation is produced.

One very promising candidate for the origin of the magnetic field
which results in afterglow emission comes from a relativistic version
of the well known two-stream Weibel instability in a plasma. In the
rest frame of a relativistic shock, the Weibel instability amplifies
any existing magnetic field perturbations by growing current filaments
out of instreaming electrons.  The amplification of the current
filaments is fed by the kinetic energy of the inflowing material (the
ISM in the case of GRBs) and does not saturate until all of the energy
in the particle distribution function anisotropy is converted into
magnetic field energy. Medvedev \& Loeb (1999) have shown that the
field is generated in the plane of the shock and has a coherence
length of order the relativistic skin depth of the shock.  Further
numerical studies (Noshed et al. 2003; Silva et al. 2003; Fredriksen
et al. 2004) have confirmed the existence and growth of this
instability.  For a detailed discussion of the Weibel instability see
Medvedev \& Loeb (1999). The major factor that makes magnetic fields
generated via the Weibel instability unique is that the correlation
length of the fields is less than a Larmor radius and, as a result,
the radiation generated by electrons in Weibel fields is very
different from the one generated by uniform large scale fields which
result in synchrotron radiation. Additionally, the field is generated
in the plane of the shock and is then transported downstream as the
shock propagates into the ISM. This continual generation of field is
an attractive solution to the problem of how field is either generated
or carried far enough downstream to produce the radiation observed in
afterglows.

In this paper we adopt the model introduced by Fleishman (2006b) and
Medvedev (2006) to describe the magnetic field in Fourier space. 
We emphasize that such a parametrization of the field by no means
incorporates all the detail and phenomenology of the fields observed
in PIC simulations of collisionless shocks. However, it has the great
advantage to be analytically tractable and relatively simple. Direct
computations of radiation spectra can be coupled to simulations
(e.g. Hededal 2005) but they are time consuming and, albeit with a
very high accuracy, provide spectral calculations valid only for that
particular simulation.  We further note that we have explicitly
excluded the presence of electric fields in our work.  Hededal (2005)
found that the electric fields are generated in PIC simulations and
have $\sim 10\%$ of the energy associated to the magnetic fields. He
also concluded that the primary effect of electric fields was to
flatten the low frequency slope.  We expect the electric fields to
play a secondary role in the generation of the emergent spectrum at
the wavelengths we consider and as such we have simplified the
analytics and numerics by excluding it.

During the initial stage of the Weibel instability, the field is
generated in the plane of the shock, with $k_\perp \sim$ skin depth
and $k_\parallel=0$. At this point, the parallel and transverse
spectra are decoupled.  Soon after saturation, secondary instabilities
develop in the parallel direction thus making $k_\parallel>0$. Since
the instability develops in the parallel direction only, there is no
coupling to the transverse dynamics either, at this time, so the
spectra are independent. Some coupling of the transverse and parallel
dynamics begins to develop at late times, when the filaments start to
twist, interact with each other and merge. At this point, we shall
expect that our assumption of the separability of the field spectrum
becomes an approximation. This is important especially in the
afterglow phase, since it has been shown that afterglow radiation has
to be produced in the whole blast wave and not only behind the shock
(Rossi \& Rees 2003). However, as long as the current filaments are
not completely intertwined to form a statistically homogeneous state,
one shall expect some degree of statistical independence of the field
spectra to remain.  In our description, the correlation tensor for the
field is given by

\beq
K_{\alpha \beta}({\bf k})=4 \pi C(\delta_{\alpha \beta}-n_\alpha n_\beta)
f_z(k_\|) f_{xy}(k_\perp).
\label{eq:B}
\eeq
\noindent

We assume that the field in the plane of the shock and the field in
the direction of propagation can be factored from each other where
$f_{xy}(k_\perp$ =$(k_x^2+k_y^2)^{1/2})$ describes the field in the
plane of the shock, $f_z(k_\|$ = $k_z)$ describes the field in the
direction of the shocks propagation, and $C$ is a normalization
constant.  The functions used to describe the field distribution are
taken from Medvedev (2006) in order to facilitate a direct comparison
with the results therein and are given by

\beq
f_z(k_\|)=\frac{k_\|^{2\alpha_1}}{(\kappa_\|^2+k_\|^2)^{\beta_1}}, \quad 
\label{fpar}
\eeq
and
\beq
f_{xy}(k_\bot)=\frac{k_\perp^{2\alpha_2}}{(\kappa_\perp^2+k_\perp^2)^{\beta_2}}
\label{fperpx}
\eeq 
 
\noindent
where $\alpha_{1,2}$, $\beta_{1,2}$, and $\kappa_{\perp,\|}$ are
parameters used to fit the spectrum to numerical results. In this
paper we have chosen $\alpha_1=\alpha_2$, $\beta_1=\beta_2$, and
$\kappa_\perp=\kappa_\|=k_B$ of the local field.  In general $\kappa$
is a parameter determined by local quantities (Medvedev et al. 2005)
but is treated as a constant in this paper as it does not
significantly alter the shape of the spectrum in the regimes under
consideration.

Finally, $C$ is fixed by the requirement that
$\int{}K_{\alpha\alpha}({\bf k})d{\bf k}=\left<B^2\right>$ where
$\left<B^2\right>$ is the mean square value of the local magnetic
field .  This convention for $C$ results in a normalization of

\beq 
C=\frac{\left<B^2\right>}{\int\!\! f_z(k_\|)k_r f_{r}(k_r)dk_\|dk _r}
\label{C}
\eeq

\noindent
where we have switched from Cartesian to cylindrical coordinates to
simplify the integration.  Unlike previous works, we have chosen to
normalize the correlation tensor by the mean square value of the
magnetic field in order to obtain absolute values for the luminosity
of the afterglow and prompt emission.

\subsection{Emission - $j_{\omega}$}
We now calculate the emissivity of an ensemble of isotropically
distributed electrons in the jitter regime. The emissivity of a
power-law distributed electrons, $N\propto\gamma^{-p}$ with a sharp
low-energy cutoff $\gamma\le\gamma_{\rm min}$ has been calculated for
the simple one-dimensional jitter model in the original paper by
Medvedev (2000).  In two separate, fully three-dimensional treatments
of jitter radiation (Fleishman 2006b; Medvedev 2006) the single
electron spectral power has been calculated.  The ensemble emissivity
is computed as a convolution of the single electron spectral power
with the electron distribution.  We present here, for the first time,
the total radiation emitted by a distribution of electrons using the
coefficients which return the true spectrum as a function of local
conditions.  Other than the powers on the correlation tensor
describing the magnetic field (to which the radiation spectrum is
relatively insensitive) every attempt has been made to keep the number
of free parameters to a minimum.

The formula used to describe the angle averaged radiation emitted by a
single, relativistic particle traveling through small scale magnetic
fields is given (neglecting plasma dispersion which change the low
frequency results to $\propto \omega^2$) by (Landau \& Lifshitz, 1971,
section 77 p. 215)

\beq
\frac{dW}{d\omega}=\frac{e^2\omega}{2\pi
c^3}\int_{\omega/2\gamma^2}^\infty \frac{\left|{\bf
w}_{\omega'}\right|^2}{\omega'^2}
\left(1-\frac{\omega}{\omega'\gamma^2}+\frac{\omega^2}{2\omega'^2\gamma^4}
\right)\,d\omega',
\label{dW/dw}
\eeq

\noindent
where $\gamma$ is the Lorentz factor of the particle. Due to
approximations used in deriving equation~(\ref{dW/dw}), this equation
is only valid when the ratio of a particle angular deflection due to
magnetic field fluctuations ($\alpha$) to its relativistic beaming
angle ($\Delta\theta\sim1/\gamma$) is much less than one. This
approximation is expounded upon later in the paper. In keeping with
Medvedev (2000) we define this ratio as

\beq
\delta\equiv\frac{\gamma}{k_B\rho_e}\sim\frac{\alpha}{\Delta\theta}
\eeq

\noindent
where $\rho_e$ is the Larmor radius of an electron. In practice the
value of $\delta$ calculated, for our parameters, is $\lsim1$.  The
value of $\delta$ enters into the integration limits in
Eqs.~\ref{form1} and~\ref{form2} below, and setting it equal to zero
there has a small effect.  The effect of this assumption can be
seen in Figure 4 of Medvedev (2000). Intermediate values of $\delta$
produce an upturn in the spectrum just below the peak that disappears
when $\delta\lsim0.3$. The shift in the peak frequency and the
normalization, due to the change in $\delta$, are not lost in our
computation.

The term $|{\bf w}_{\omega'}|^2$ in equation~(\ref{dW/dw}) is the
square of the Fourier transform of the acceleration field due to the
Lorentz forces.  Here we replace it with a volume averaged
$\langle|{\bf w}_{\omega'}|^2\rangle$ by assuming a statistically
homogeneous turbulence.  The derivation of this term is left for the
appendix and has been included for the sake of completeness (the
interested reader is referred to Fleishman 2006b for the full details)
and we state here the result assuming the fields described above

\beq
\langle|{\bf w}_{\omega'}|^2\rangle =({\frac{e}{\gamma m_e}})^2 
\frac{CT}{2\pi}(1+\cos^2\Theta')I(\Theta'),
\label{w-main}
\eeq

\noindent
where I($\Theta'$) is given by 

\beq 
I(\Theta')=\int\!\! f_z(k_\|) f_{xy}(k_\perp)
\delta(\omega'+{\bf k\cdot v})\,d^3 k,
\label{integrand}
\eeq
\noindent
where $\Theta'$ is the angle between the particle velocity and the
observer in the comoving frame, $C$ is given by equation~(\ref{C}),
$T$ is the period for an electron traveling in Weibel fields, and
$f_z(k_\|)$ and $f_{xy}(k_\perp)$ are given by equations~(\ref{fpar})
and~(\ref{fperpx}), respectively. To evaluate the integral in
equation~(\ref{integrand}) it is necessary to specify both the limits of
integration and the geometry of the problem.  The limits come from the
nature of the magnetic fields generated by the Weibel instability.
The Fourier component associated with the fastest growing mode in the
Weibel instability is

\beq
k_{Weibel}=\frac{4\gamma_{\rm shock}\omega_{pe}}{2^{1/4}\bar\gamma_e^{1/2}c}.
\label{kBe}
\eeq

The factor of $4\gamma_{shock}$ comes from the shock compression and
$\omega_{pe}^2=\frac{4\pi e^2 n_{Ext}}{m_e}$ is the plasma frequency
of the pre-shocked material (which, in the case of afterglows,
corresponds to the ISM or wind density profile). While modes are
initially compressed only perpendicularly to the shock plane, we make
the assumption that the spatial scales are mixed by turbulence and set
the inverse length scales, $k_\|$, $k_\perp$ = $k_{Weibel}$.

The geometry of the system is defined as follows: The shock is
propagating in the z direction and lies in the x-y plane.  A particle
in the shock has a velocity vector given by ${\bf k}={\bf \hat
x}k\sin\Theta' + {\bf \hat z}k\cos\Theta'$, this gives us ${\bf k\cdot
v}=k_x v\sin\Theta'+k_z v\cos\Theta'$. Medvedev (2006) presented three
separate forms for $\langle|{\bf w}_{\omega'}|^2\rangle$,
corresponding to a shock viewed at 0 degrees, at $\pi/2$, and in
between these extremes.  We chose to evaluate
equation~(\ref{integrand}) somewhat differently by using the
properties of the delta function to integrate the correlation function
in two distinct ways.  By doing this we can match the asymptotic forms
with two functions as opposed to three and do not suffer from
numerical errors. The two forms are necessary to avoid the
introduction of infinities as $\Theta$ approaches $0$ and $\pi /2$
degrees.  The two forms we use are

\beq 
I_1(\Theta')=\int\frac{1}{|v\cos\Theta'|}f_z\!\left(
\frac{\omega'/v}{\cos\Theta'} + k_x\tan\Theta' \right) f_{x\bar y}(k_x)
dk_x,
\label{form1}
\eeq
and
\beq
I_2(\Theta')=\int\frac{1}{|v\sin\Theta'|}f_z(k_z)f_{x\bar y}\!\left( 
\frac{\omega'/v}{\sin\Theta'} + \frac{k_z}{\tan\Theta'} \right)dk_z,
\label{form2}
\eeq

\noindent
where the bar in $f_{x\bar y}$ denotes an integration over $k_y$. In
Fourier space, the modes that generate the field lie within a
spherical annulus with radius $k_\perp$ and $k_\|$ $\in$ $[\delta
k_{Weibel}, k_{Weibel}]$ and this sets the limits for the integrals in
equations~(\ref{form1}) and~(\ref{form2}).  Our magnetic field model
assumes that fluctuations in the magnetic field are not generated on
physical scales larger than $\sim (\delta k_{Weibel})^{-1}$ and
therefore these scales are excluded from the integration region.  In
practice, these large scales contribute little to the overall spectrum
for $\delta \ll 1$ and this is seen by the lack of any change when we replace
$\delta k_{Weibel}$ by zero in our code.

Choosing $v\sim c$ and switching between forms for $\Theta'\sim$
$\frac{\pi}{4}$ we are able to reproduce the results of Medvedev
(2006).  Combining equations~(\ref{fpar}), (\ref{fperpx}), (\ref{C}),
(\ref{dW/dw}), (\ref{w-main}), (\ref{form1}), and~(\ref{form2}) and
multiplying the result by $\frac{1}{T}$ gives us 
\beq
P(\omega)=\frac{1}{T}\frac{dW}{d\omega}:
\label{pw}
\eeq 
\noindent
the total energy per unit frequency and unit time, emitted by a single
electron in a magnetic field with the configuration specified by
Eq.~\ref{eq:B}.  Unlike the case of synchrotron, where it is possible
to precisely define an orbital period, the random, small scale nature
of Weibel turbulence requires a more arbitrary choice for $T$. We
follow Medvedev (2006) and Fleishman (2006b) which includes the value
of the period in the magnetic field derivation (see
equation~(\ref{w-main})).  By using this method, $T$ simply disappears
from the final form for $P$($\omega$).

The final step required to calculate the emission from a population of
electrons is to choose the form for the distribution.  It is standard
in GRBs and other sources of non-thermal radiation to assume that the
electrons follow a power law distribution 
\beq 
n'(\gamma)=K\gamma^{-p}
\label{powerlaw}
\eeq
\noindent
where $\gamma \in [\gamma_{min},\infty]$, $K = (p -
1)n'\gamma^{p-1}_{min}$ and 
\beq
\gamma_{min} = \frac{p -2}{p - 1}
\frac{\epsilon_e e'}{n'm_ec^2}
\eeq
The primed quantities $n'$ and $e'$ refer to the comoving number and
energy densities and $\epsilon_e$ is the fraction of the thermal
energy in the electrons.

Combining equations~(\ref{powerlaw}) and~(\ref{pw}) and integrating
from $\gamma_{min}$ to $\infty$ yields the final result 
\beq 
P_{\rm tot}(\omega)=
\int_{\gamma_{min}}^\infty n'(\gamma)P(\omega)d\gamma.
\label{pensemble}
\eeq

It is this equation that we have numerically solved for several values
of $\Theta'$. In order to compute $j_{\omega}$ we make several
assumptions.  First, we assume that there is no scattering.  Second we
assume an isotropic electron distribution.  Third, we assume that for
a given point on a surface, whether it is a plane parallel slab or a
spherically expanding shell, we only see the radiation due to the
comoving angle aligned to the local bulk velocity.  This third
assumption is physically equivalent to assuming we see only
'flashlights' from different emitting regions.  Finally we assume that
the medium is moving ultra-relativistically and is beamed into a
narrow cone of angular width $1/\gamma$.  Using these assumptions we
make the simplifying approximation that $j_\omega
=P_{tot}(\omega)\delta(\Omega-\Omega(\hat{r}))$.  Bearing in mind that
we have so far calculated all quantities in the comoving frame, this
form for the emissivity introduces an additional factor of
$\gamma^2(1+\beta\cos\Theta')^2$ when transforming to the observer's
frame due to the transform of the solid angle (Granot, Piran, \& Sari
1999a; Rybicki \& Lightman 1979).
 
In order to check our results we have performed the following tests.
First, it can be shown analytically (for $\beta_{1,2} = 0$) that
combining the $\Theta'$ = 0 form of $\langle|{\bf
w}_{\omega'}|^2\rangle$ with $P=(2e^2\gamma^4/3c^3)(w^2_\perp +
\gamma^2w^2_\|)$ (equation~4.92 of Rybicki \& Lightman 1979)
Eq.~\ref{pw} reduces to
\beq 
dW/dt = (2/3) r_e^2 c\gamma^2\left<B^2\right>, 
\eeq
\noindent
which is equivalent to the case of synchrotron radiation. We have
confirmed that our procedure numerically returns the same value by
integrating equation~(\ref{dW/dw}). Setting $\beta_{1,2} \neq 0$ in
our code does not change this result. Second, we replaced
$P$($\omega$) in equation~(\ref{pensemble}) with the the form for
synchrotron radiation and verified that our technique yielded results
which were in agreement with the analytic form for $P_{\rm
tot,Synch}(\omega)$ given by equation~6.36 in Rybicki \& Lightman
(1979).

\subsection{Absorption - $\alpha_{\omega}$}

Given the single particle emissivity, $P(\nu)$, and the particle
energy distribution, the calculation of the jitter self-absorption is
straightforward.  We have analytically calculated it (Medvedev, et
al. 2007) in the regimes when the self-absorption frequency is above
and below the jitter peak frequency. For numerical implementation, the
self-absorption coefficient $\alpha_{\omega}$ can be derived using
equation~6.50 from Rybicki \& Lightman (1979) as follows: 
\beq
\alpha_{\nu} = -\frac{c^2}{8\pi\nu^2}\int P(\nu,E)E^2
\frac{\partial}{\partial E}[\frac{N(E)}{E^2}]dE 
\eeq 

Using $P(\nu,E)=2\pi P(\omega)$, $E=\gamma m_e c^2$, $N(E) =
\frac{d\gamma}{dE} n(\gamma)$, and $\omega = 2\pi\nu$ where
$n(\gamma)$ and p($\omega$)are given by equations~(\ref{pw})
and~(\ref{powerlaw}), quickly yields 
\beq 
\alpha_{\omega}=
\frac{(p+2)\pi^2 k}{m_e \omega^2}\int_{\gamma_{min}}^\infty
\gamma^{-(p+1)}P(\omega)d\gamma.
\label{absorb}
\eeq

Effectively, once the numerical problem of solving
equation~(\ref{pensemble}) has been solved, the only modification
necessary to obtain the absorption coefficient is a change in the
power on gamma and a change in the normalization constant.  In using
this form for the absorption coefficient we have assumed that the
electron distribution is locally isotropic, that no scattering occurs,
and that the local absorption is dominated by electrons moving with
only one angle with respect to the observers line of sight.  For a
further discussion of the absorption coefficient see Medvedev et
al. (2007).

\section{Limitations of Validity} 

Equation~(\ref{dW/dw}) is derived using the perturbation theory
(Landau \& Lifshitz 1971) and, consequently, has a limited range of
applicability.  Jitter radiation explicitly assumes that the coherence
length of the emission is small enough that the approximation of
rectilinear motion is valid (Medvedev 2000 \& Fleishman 2006b). The
coherence length $\lambda_c$ is proportional to $1/\omega$ and
clearly increases as the frequency decreases.  Fleishman (2006)
describes that, at lower frequencies, a particle trajectory random
walks due to repeated scatterings off magnetic field
inhomogeneities. This scattering results in a modification of the
emitted radiation in the region of the spectrum just before where the
plasma dispersion dominates. The non-perturbative version of the
theory of radiation emitted in small scale field (Toptygin \&
Fleishman 1987, Fleishman 2006a) accounts for this effect and finds a
$P(\omega)\propto\omega^{\frac{1}{2}}$ regime in this region.  At even
lower frequencies, $\omega\la\omega_{pi}$, the spectrum follows
$P(\omega)\propto\omega^{2}$ due to plasma dispersion. This region is
at photon energies of a fraction of an eV or less in prompt GRBs and
at frequencies below $\sim 1$~MHz in afterglows, assuming standard GRB
parameters.

The perturbative form of the theory fully accounts for the spectrum
above and below (down to the plasma dispersion regime) the
$P(\omega)\propto\omega^{\frac{1}{2}}$ region.  For the purposes of
this work it can be shown that the region in frequency (approximately)
given by $ \omega < \epsilon_B \omega_{jm}$ (where $\omega_{jm}$ is
the peak frequency of jitter radiation for a given minimum Lorentz
factor) is affected by scattering and therefore the spectrum obtained
with the methods used in this paper is only approximate. This
limitation does not pose a problem for this work. It can be shown that
$\delta \approx 36 \sqrt{\epsilon_B}$ (Medvedev et al. 2007) which
requires we assume $\epsilon_B < 10^{-3}$ and, more realistically, in
the range of $10^{-6} - 10^{-4}$.  Even for a value of $\epsilon_B =
10^{-2}$ the cutoff frequency $\epsilon_B \omega_{jm}$ corresponds to
an energy at the edge of the BATSE cutoff window of 20 keV.

The only case in which the limitations of the perturbation approach
may be relevant is the modeling of the afterglow spectrum.  Even in
such a case, however, the region in question generally lies entirely
within or close to the optically thick portion of the spectrum, in
which case the spectrum simplifies to the source function.  In cases
where the spectrum generated by our perturbative approach may be
incorrect we will note it in the individual results. 


\section{Results}

We here consider a single internal shock within the prompt phase of a
GRB.  The material has a pre-shock density of
$9.25\times10^{-12}$~g/cm$^3$, a shock Lorentz factor of $\gamma_{int}
= 2$, $\epsilon_e=0.2$, and a magnetic equipartition
fraction\footnote{\bf We note that there is no experimental value for
the value of $\epsilon_B$ during the prompt phase.}
$\epsilon_B=10^{-5}$.  The spectral index of the electrons is $p=2.5$.
The thickness of the shell is $3\times10^{6}$~cm in the observer frame
and the material is moving with a bulk Lorentz factor of $100$. We
consider a shell radiating at $R=3.2\times10^{14}$~cm. This setup
corresponds to an isotropic energy of $5\times10^{52}$~ergs in the
shell.  

These values are chosen to give a peak frequency of about $200$~keV in
the observer frame.  The parameters describing the magnetic field
distribution do not affect the spectral characteristics as long as
$\alpha_{1,2}>0.5$ and $\beta_{1,2}>\alpha_{1,2}+p/2$. To be
consistent with Medvedev (2006) we set $\alpha_{1,2}=2.$ and
$\beta_{1,2}=10$.  It is important to keep in mind that the emission
angles correspond to co-moving angles.  The results plotted are the
observed quantities as a function of co-moving emission angle.

\begin{figure*}
\psfig{file=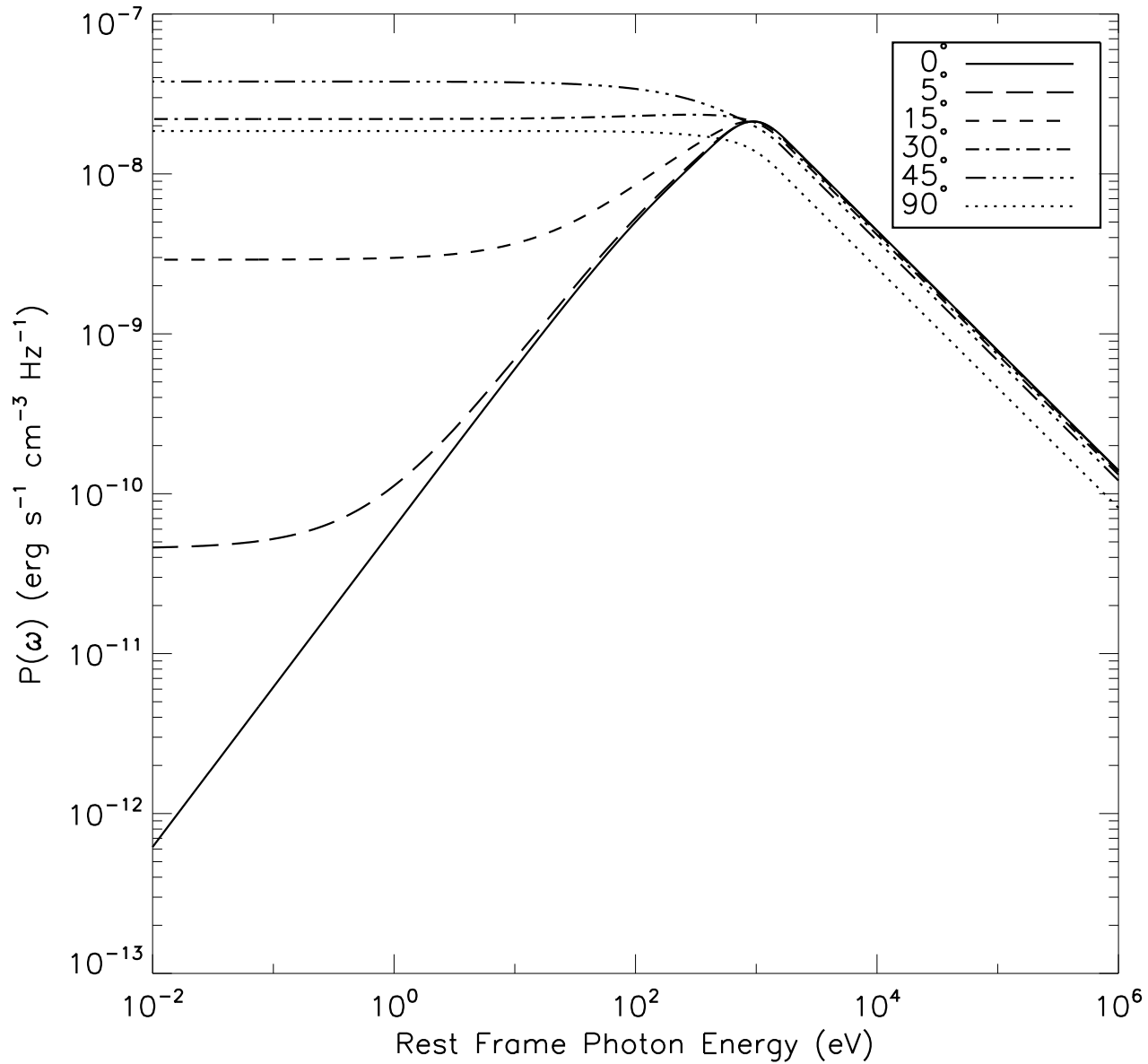}
\caption{{Jitter emissivity vs. frequency at different angles between
the bulk velocity vector and the line of sight.  The angles plotted
are $\theta'=0\degr$ (solid line), $5\degr$ (long-dashed line),
$15\degr$ (dashed line), $30\degr$ (dot-dash line), $45\degr$ (3
dot-dash), and $90\degr$ (dotted line). All the angles are measured in
the comoving frame.}
\label{fig:1}}
\end{figure*}

\begin{figure*}
\psfig{file=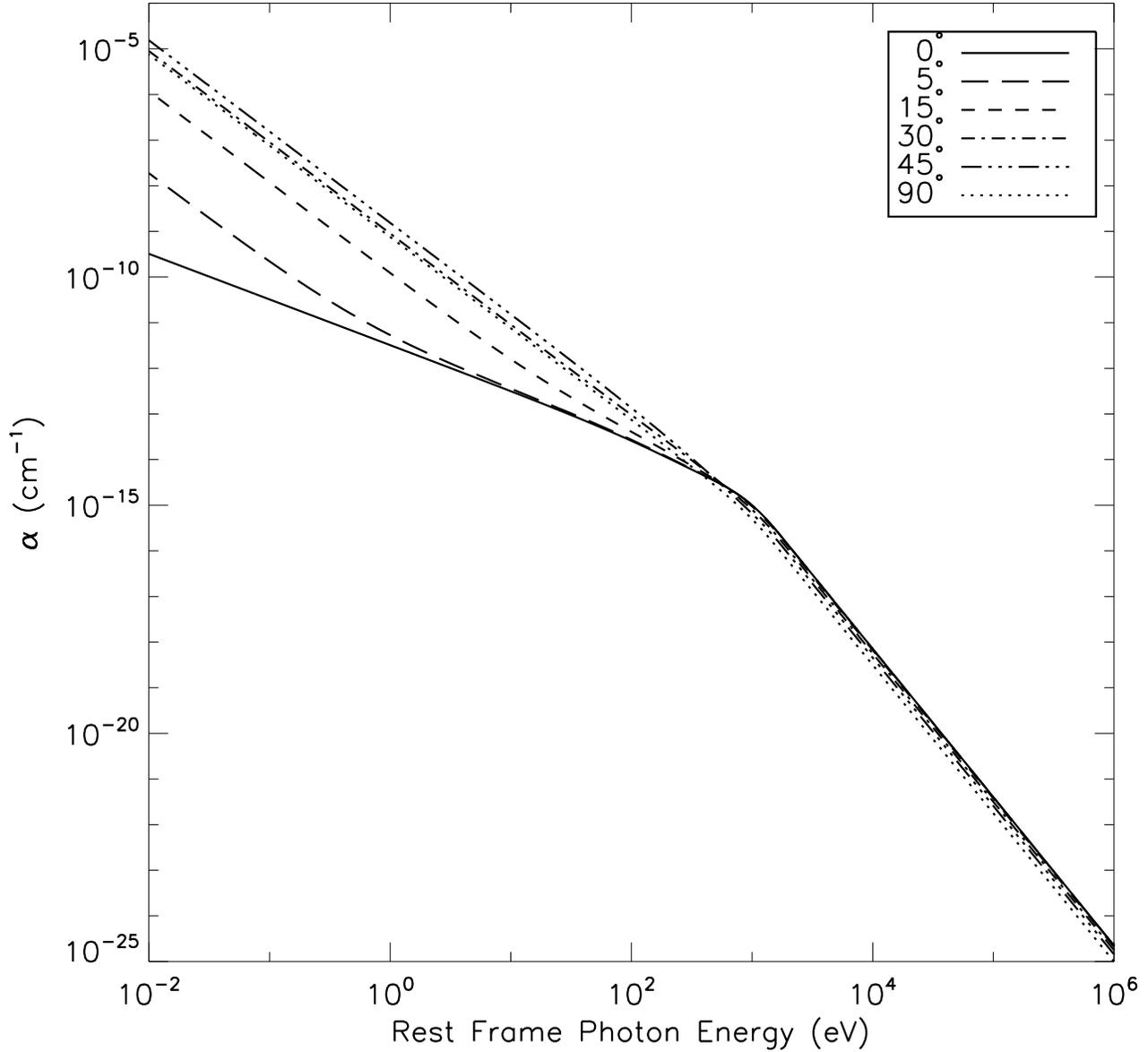}
\caption{{Absorption coefficient $\alpha$ vs. frequency at different
angles between the bulk velocity vector and the line of sight.  The
angles plotted are $\theta'=0\degr$ (solid line), $5\degr$
(long-dashed line), $15\degr$ (dashed line), $30\degr$ (dot-dash
line), $45\degr$ (3 dot-dash), and $90\degr$ (dotted line).}
\label{fig:2}}
\end{figure*}

Figure~\ref{fig:1} shows the emissivity vs. frequency for different
emission angles ($\theta'$). For $\theta'=0$ the emission has a slope
$\alpha_\nu=1$ at low frequencies and then turns over to a slope
$\alpha_\nu=-(p-1)/2=-0.75$, set by the electron power index.  As the
angle increases, the low frequency slope near the peak decreases, and
at much lower frequencies the slope flattens to $\alpha_\nu=0$.
Beyond an angle of $30\degr$, there is no longer a peak in the
spectrum, just a transition from a constant ($\alpha_\nu=0$) to
$\alpha_\nu=-0.75$.  The value of the low frequency constant changes
with angle, peaking at $45\degr$.  At high frequencies, the slope is
always the same but the emissivity at a given frequency decreases with
angle. These results are in full agreement with the single electron
spectra of Medvedev (2006).  Figure~\ref{fig:2} shows the absorption
coefficient $\alpha$ for the same setup.  The absorption coefficient
is proportional to $P(\omega)\times\omega^{-2}$ below the peak
frequency and to $P(\omega)\times\omega^{-2.5}$ above the peak
frequency.

Below frequencies of $\sim3\times10^{-3}$~eV our perturbative approach
is no longer valid.  A non-perturbative approach would be needed to
properly calculate radiation at lower frequencies (Fleishman 2006b).
Therefore, frequencies in this regime are not considered in this
paper.

\begin{figure*}
\psfig{file=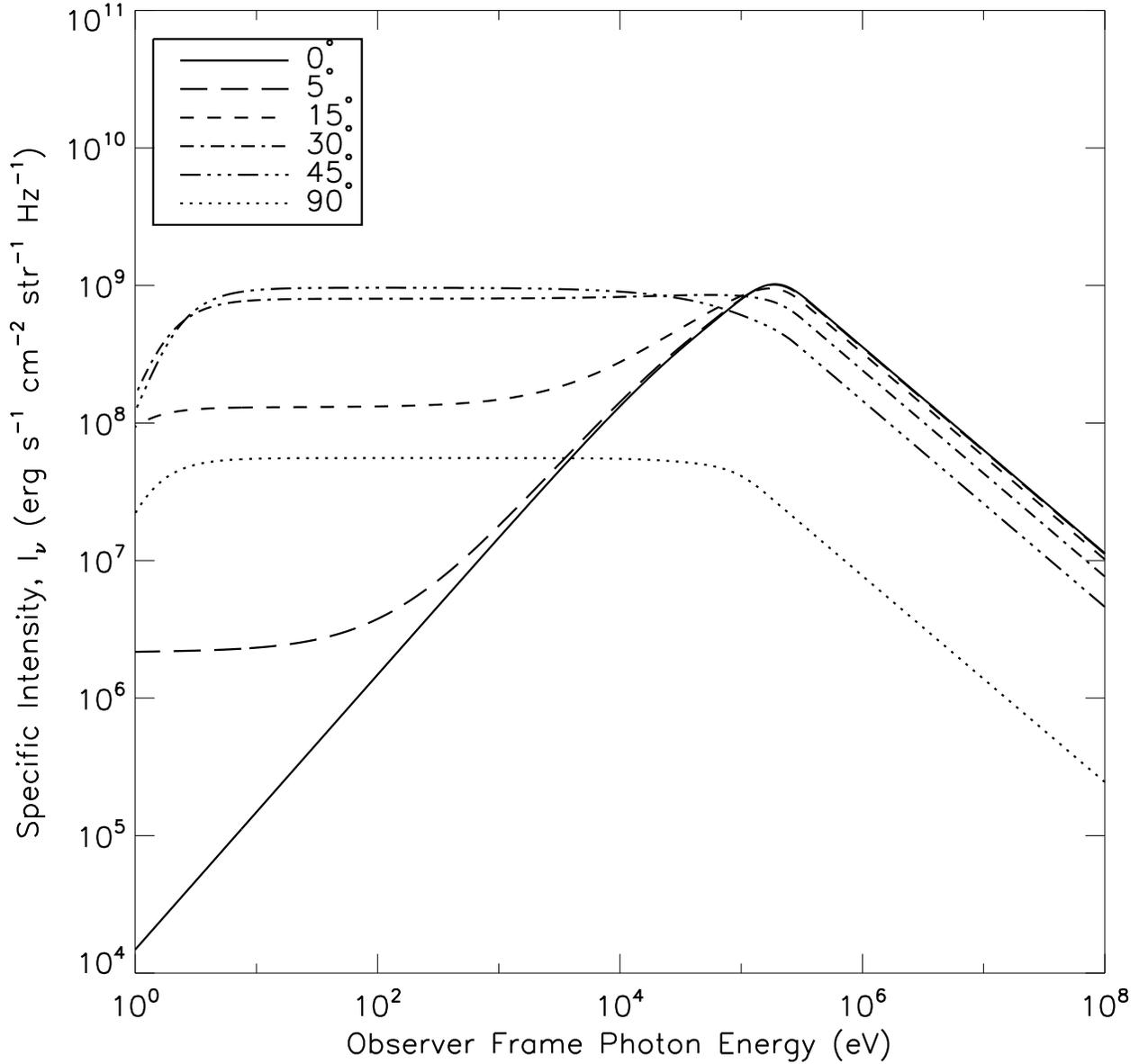}
\caption{{Jitter emission of a single relativistic shell as seen from
different angles (angles measured in the comoving frame).  The angles
plotted are $\theta'=0\degr$ (solid line), $5\degr$ (long-dashed
line), $15\degr$ (dashed line), $30\degr$ (dot-dash line), $45\degr$
(3 dot-dash), and $90\degr$ (dotted line).  Jitter radiation peaks at
about $190$~keV.}
\label{fig:3}}
\end{figure*}

\begin{figure*}
\psfig{file=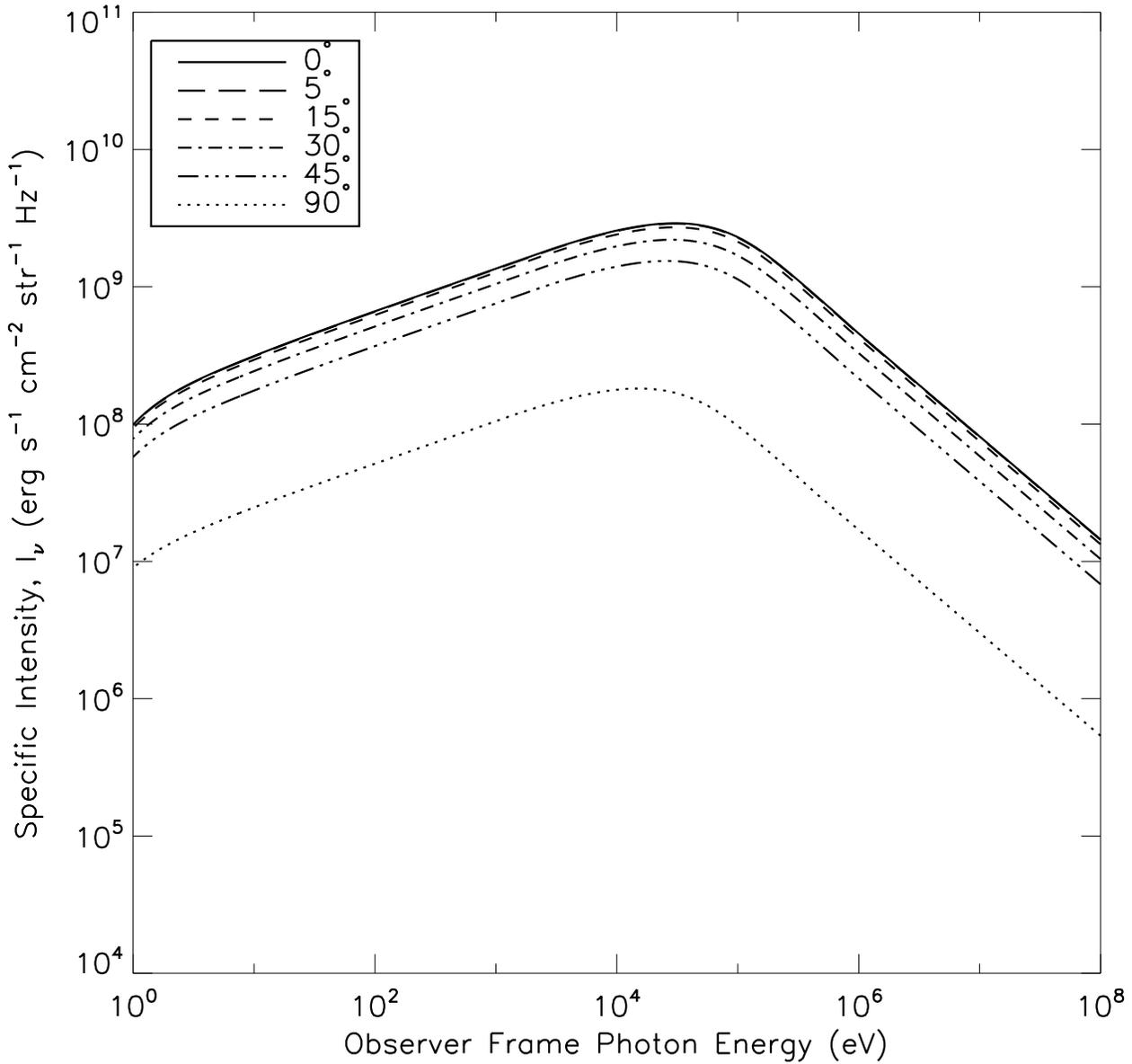}
\caption{{Synchrotron emission for a single relativistic shell at
different angles (analogous to Fig.~\ref{fig:3}).  The angles plotted
are $\Theta'=0\degr$ (solid line), $5\degr$ (long-dashed line),
$15\degr$ (dashed line), $30\degr$ (dot-dash line), $45\degr$ (3
dot-dash), and $90\degr$ (dotted line).  Synchrotron radiation peaks
at about $31$~keV}
\label{fig:4}}
\end{figure*}

With $P(\omega)$ and $\alpha$ we can solve the 1D radiative transfer
equation to find the specific intensity of the shell at different
angles.  Assuming the observable portion of the shell is spherical and
that the amount of time any piece of the shell is radiating is short
compared to the time it takes the shell to become visible (i.e., we
assume the shell is infinitely thin), the specific intensity at a
given angle is proportional to the total emission at the time that
angle comes into view.  Figure~\ref{fig:3} shows the integrated
specific intensity at different angles for jitter radiation.
Figure~\ref{fig:4} shows synchrotron radiation for identical
conditions and angles.

The synchrotron radiation spectrum shows no change in the shape of the
spectrum.  The peak frequency decreases by about a factor of $2$ and
the power decreases by a factor of $\sim16$ between $0\degr$ and
$90\degr$ due to relativistic effects.  Jitter radiation, on the other
hand, shows a large change in spectral index and total emission below
the peak frequency and a decrease in emission of about a factor of
$\sim27.5$ with angle above the peak frequency.  Note that the source
function for jitter and synchrotron are the same for identical
conditions.  Self absorption becomes important at a few 10s of eV in
this example.

\begin{figure*}
\psfig{file=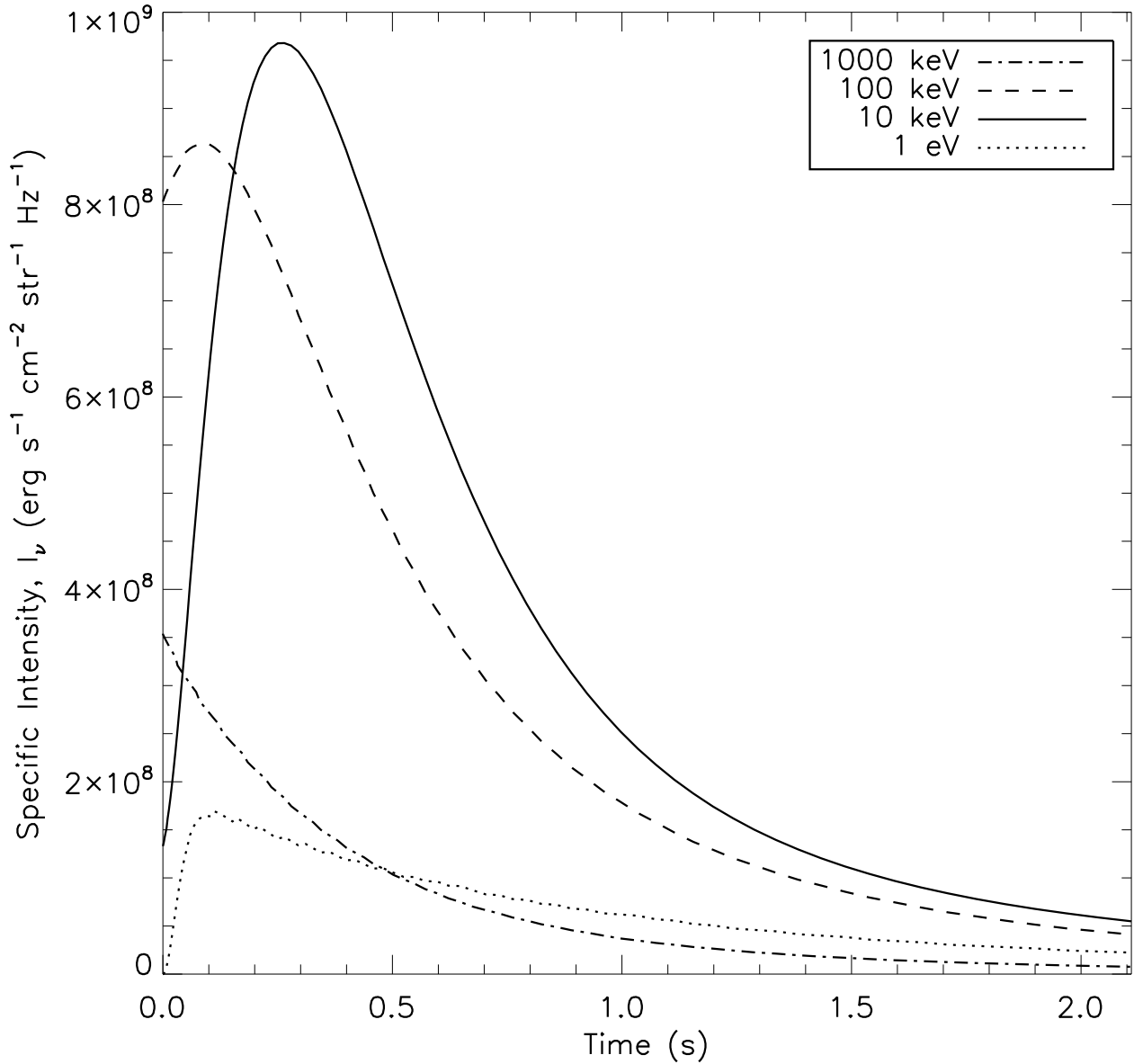}
\caption{{Jitter radiation light curves for a single thin relativistic
shell at 1~MeV (dot-dash line), 100~keV (dashed line), 10~keV
(solid line) and 1~eV (dotted line).}
\label{fig:5}}
\end{figure*}

The time at which different angles come into view is found from the
geometric time delay to be $(1-\cos\theta) \times r/c$ where $\theta$
is the emission angle in the observer frame, $r$ is the emission
radius ($10^{14}$~cm in this example) and $c$ is the speed of light.
Figure~\ref{fig:5} shows intrinsic brightness vs. time at $10$, $100$,
and $1000$~keV and $1$~eV ($124$~nm).  At $1000$~keV, above the
initial peak frequency, the emission decreases with time by a factor
of about $46$ between $t=0$ and $t=2.1$.  At $100$~keV, just below the
initial peak frequency, the emission increases slightly for $0.1$~s
and then decreases.  At $10$~keV the emission increases by a factor of
$7$ in $0.25$~s, corresponding to $40\degr$, and then decreases.  The
initial amount of increase is larger at lower energies.  At $1$~eV,
there is an initial sharp increase. Very early on, however, the
emission becomes optically thick and begins a shallow decrease (by a
factor of only $\sim7$ with time). This behavior has important
implications for the interpretation of spectral lags and the optical
emission during the prompt phase. As is clear from Fig.~\ref{fig:5},
the optical and high energy bands can show very little correlation,
even if they come from the same electrons and the same emission
mechanism. Lack of correlation was used as an argument in favor of a
reverse shock origin of the optical flash in GRB~990123 (Akerlof et
al. 1999, Sari et al. 1999). The flattening of the spectra in the
X-ray band could also explain the soft X-ray excesses detected in some
events (Preece at al. 1996; Strohmayer et al. 1998).

\begin{figure*}
\psfig{file=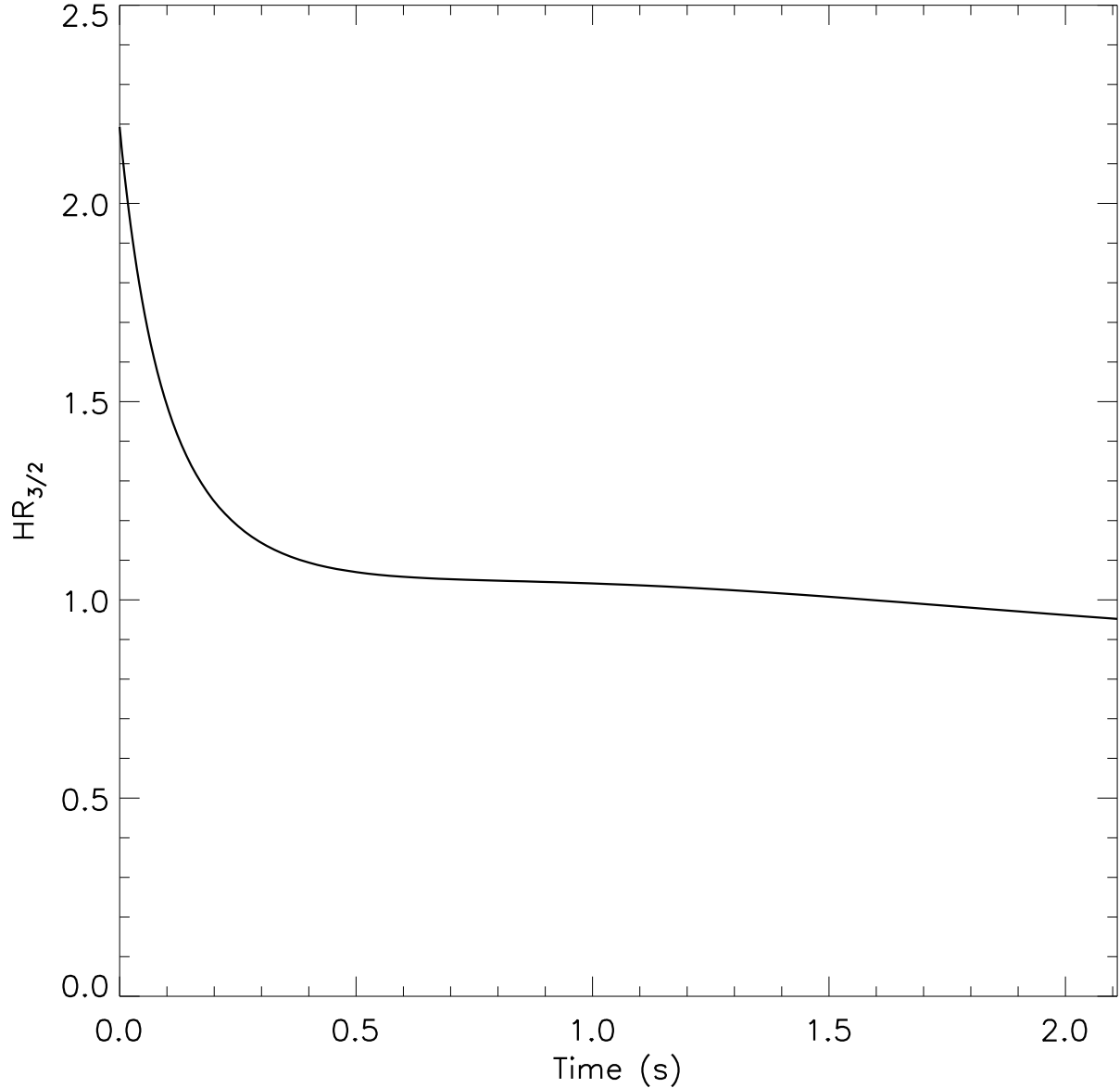}
\caption{{Simulated BASTE 3:2 channel hardness ratio vs. time.
HR$_{3/2}$ is defined as the ratio of the number of photons between
$110$ and $320$~keV to the number of photons between $55$ and
$110$~keV. }
\label{fig:6}}
\end{figure*}

Figure~\ref{fig:6} shows the ratio of the total number of photons
between $110$ and $320$~keV to the number of photons between $55$ and
$110$~keV vs. time.  This ratio is equivalent to the count ratio of
BATSE channels 3:2 (HR$_{3/2}$).  This plot shows a hard to soft
evolution for the first $0.4$~s (out to a $45\degr$ angle) and then a
fairly flat hardness ratio after this.

\begin{figure*}
\psfig{file=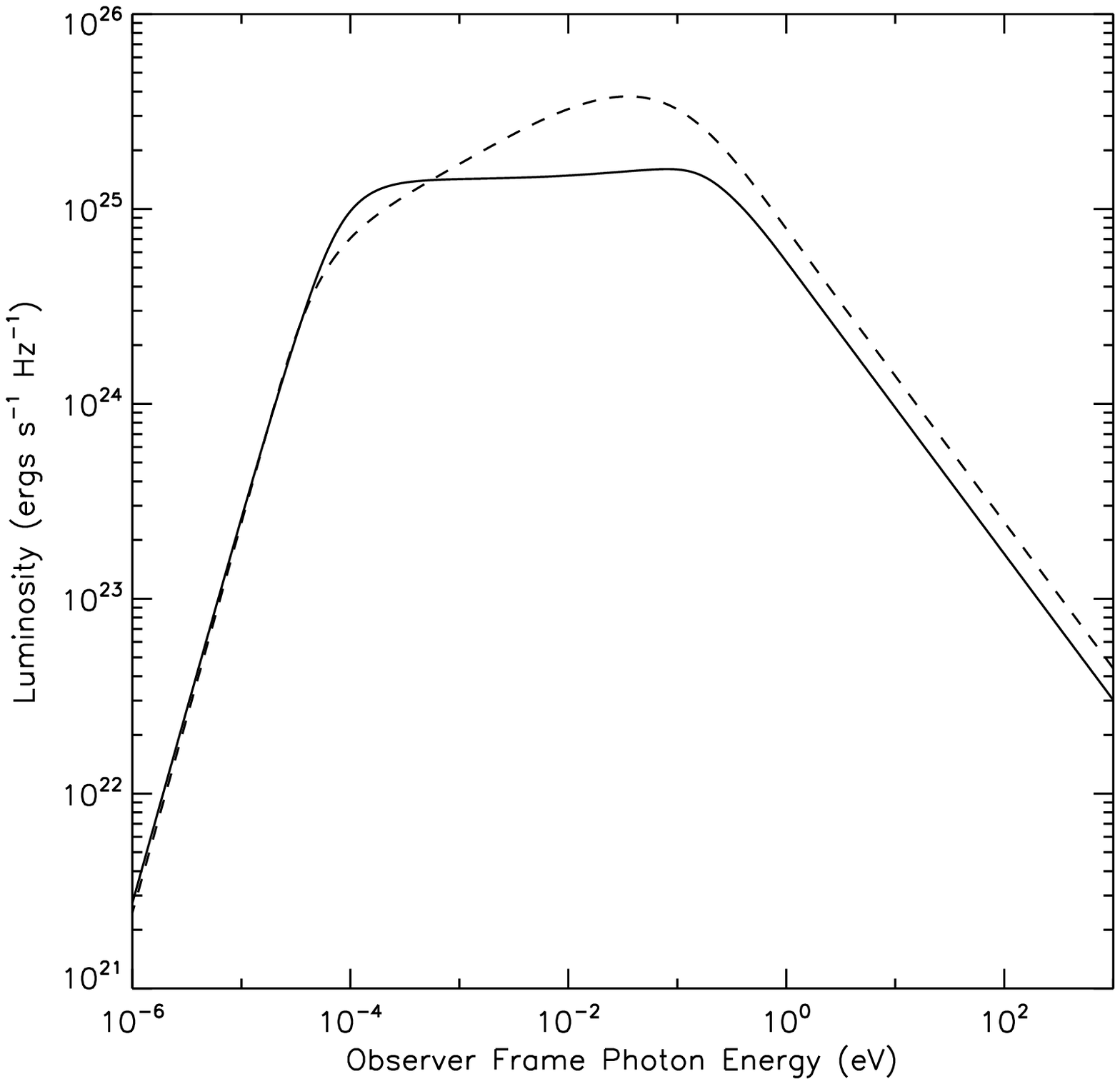}
\caption{{The total afterglow power emitted at $1000$~s for jitter
(solid line) and synchrotron (dashed line) radiation mechanisms under
identical conditions. See text for details on the computations.}
\label{fig:7}}
\end{figure*}

Let us now consider the external shock phase.  For an example
comparison of the afterglow spectrum for jitter and synchrotron
radiation we examine an isotropic explosion with an energy of
$10^{53}$~ergs expanding into a medium with a density of
$1$~particle/cm$^3$ at the observed time $t=1000$~s.  The
Blandford-McKee (Blandford \& McKee, 1976) solution for a relativistic
fireball is used to determine the density ($\sim100$~cm$^{-3}$),
internal energy ($\sim4$~erg~cm$^{-3}$)and Lorentz factor
($\Gamma\sim25$) of the material at the leading edge of the shock.  A
thin shock model is used to calculate the emission and the optical
depth.  A magnetic field equipartition parameter $\epsilon_B=10^{-5}$
and an electron equipartition parameter $\epsilon_e=0.01$ are assumed
everywhere, and a shock thickness of
$\frac{r}{2\gamma^2}\approx2.4\times10^{14}$~cm everywhere is
assumed. Figure~\ref{fig:7} shows a comparison between the emission
from jitter and synchrotron radiation under these conditions.  Below
the peak frequency, the synchrotron emission increases with frequency
as $\nu^{1/3}$, while the jitter spectrum is nearly flat.  This is due
to the fact that large angles dominate the afterglow spectrum (see
also Medvedev et al. 2007).  The synchrotron and jitter spectra
become optically thick at around $10^{-4}$~eV ($24$~GHz) and
$10^{-3}$~eV ($240$~GHz), respectively, in this example, but this is
dependent on the thickness of the afterglow material.

\section{Summary and discussion}

We have computed jitter radiation spectra for a non-thermal population
of relativistic electrons radiating in a highly non-uniform magnetic
field, including self absorption. Our results are in agreement, when
relevant, with previous computations (Fleishman 2006b; Medvedev 2006;
Medvedev, et al. 2007). Due to the complexity of the problem,
several assumption and/or simplification were made in order to obtain
a semi-analytical result for the spectra. We here review the
assumptions and simplifications made.
\begin{itemize}
\item The most important simplification we adopt is that of assuming
that the field can be factorized in the parallel and perpendicular
directions with respect to the shock propagation velocity. This
assumption is fundamental in our results as it allows for the
dependence of the low-frequency spectral slope on the viewing
angle. It is likely to be a good assumption immediately behind the
shock and become progressively worse as the ion current filaments get
twisted and/or merge, mixing the parallel and perpendicular components
of the field.
\item The strength of the magnetic field is assumed to be uniform
throughout the shocked region and the average magnetic field to be
rather small (containing only a fraction $\sim10^{-5}$ of the internal
energy of the blast wave). Such assumption is necessary to simplify
the computation and to make sure the radiation is produced in the
jitter regime rather than in a transition region. PIC simulations
(Silva et al. 2003; Fredrikssen et al. 2004; Cheng et al. 2007) find a
stronger magnetic field right behind the shock and a moderate decline
behind it. However, PIC simulations can follow the shocked material
only several hundred skin depths behind the shock, while afterglow
radiation has to be produced by the whole blast wave, in order to
avoid a strong inverse Compton component that is not observed (Rossi
\& Rees 2003).
\item The peak of the magnetic spectrum can be around the ion skin
length and this seems to invalidate the jitter condition for
leptons. However, this does not invalidate the use of jitter
radiation, because the deflection angle depends on the product of the
field correlation length and the field strength. We took this into
account: the jitter break will be above the synchrotron frequency in
the field of the same strength if $\epsilon_B\lsim10^{-3}$. Fits to
observational data of GRBs favor such low values (Panaitescu \& Kumar
2001; Table 3 of Panaitescu 2005). If the field strength is larger,
the jitter regime transforms into the wiggler regime (Attwood 2000)
and, for even stronger fields and larger correlation lengths - into
the synchrotron regime. We should also stress here that in reality the
magnetic field spectrum spans at least several decades in $k$-space,
with the highest $k$ being even higher than $\omega_{pe}/c$, even for
the electron-ion plasmas (Wiersma \& Achterberg 2004). This is also
confirmed in PIC simulations. Radiation from these fields will surely
be in the jitter regime; the low-$k$ part of the field spectrum will
produce synchrotron-like radiation, as is discussed in (Medvedev
2000).
\item Our computations are performed in the perturbative approach and
cannot be applied in some regimes. Especially relevant is the low
frequency range where the photons scatters off the magnetic field
inhomogeneities resulting in a $F(\nu)\propto\nu^{1/2}$ regime
(Fleishman 2006b).
\item Simulation of magnetic field generation in ion-electron
collisionless shocks find evidence of electric fields on top of
magnetic fields. We neglect the acceleration of the electrons due to
electric fields in our computations. A discussion of the effect of
electric fields can be found in Hededal 2005, who conclude that the
effect is that of flattening the low frequency slope.  This effect
will be important in the modeling of the prompt bursts but not in
afterglows, where the resultant spectrum will be dominated by the limb.
\item We also assume that there is no correlation between the electron
energy and the magnetic field locally. Such assumption is customary
also for synchrotron computations.
\item Finally, the limit $\delta<<1$ is assumed
\end{itemize} 

We find that depending on the orientation of the line of sight with
respect to the shock front the jitter spectrum is different. For
shocks observed head on, the spectrum is peaked, with a steep low
energy slope $F(\nu)\propto\nu$. For shocks observed edge on, the
spectrum at the left of the peak is flat down to the self absorption
frequency or to the frequency at which scattering off the magnetic
field inhomogeneities becomes important (Fleishman 2006b). We applied
these results to standard GRB cases, for the prompt and for the
afterglow emission. We find that, in addition to some of the
observations already discussed in the previous literature (Medvedev
2000, 2006), jitter radiation can explain X-ray excesses in the prompt
spectrum and the lack of correlation between optical and high energy
radiation. This is due to the fact that the optical radiation lies in
the regime where the spectral slope depends on the orientation angle,
while the high energy emission depends only on parameters that do not
evolve with time. We also find that jitter radiation explains
naturally the presence of spectral lags (Norris et al. 2001).

In the afterglow regime, the optical and X-ray spectra are very
similar to those of synchrotron radiation. At frequencies below the
peak, differences are instead clearly visible. First, the spectrum is
flat ($F(\nu)\propto\nu^0$ instead of $F(\nu)\propto\nu^{1/3}$);
second, the self absorption frequency is at higher frequencies (see
also Medvedev, et al. 2007 who also show that self absorption
frequency has a different temporal evolution in jitter and
synchrotron). Understanding the implications of those differences is
not simple and a proper fit has to be performed (Morsony et al. in
preparation). It is likely that jitter spectra fit to afterglow data
will give different results in term of the properties of the ambient
medium, since radio observations are always very important in
constraining the density of the interstellar material. A more detailed
discussions will be possible only after a formal fit has been
performed. 

\section*{Acknowledgements}

This work was supported by NSF grants AST-0407040 (JW) and AST-0307502
(BM, DL), NASA Astrophysical Theory Grants NNG04GL01G and NNX07AH08G
(JW), NNG06GI06G (BM, DL) and NNG-04GM41G (MM), Swift Guest
Investigator Program grants NNX06AB69G (BM, DL) and NNX07AJ50G (MM),
and DoE grant DE-FG02-04ER54790 (MM).  MM gratefully acknowledges
support from the Institute for Advanced Study.



\appendix
\section{Derivation of $\langle|{\bf w}_{\omega'}|^2\rangle$}

\noindent
The following derivation is directly derived from Medvedev (2006) and
Fleishman (2006b) and is included for the sake of completeness. First,
let us define a parameter $\delta$ which defines the ratio of
deflection ($\alpha$) due to Lorentz forces and beaming
($\Delta\theta$) due to relativistic effects experienced by a particle
moving with Lorentz factor $\gamma$ in a small scale, random magnetic
field with a typical correlation length, $k_B$ and Larmor radius,
$\rho_e$.

\beq
\delta\equiv\frac{\gamma}{k_B\rho_e}\sim\frac{\alpha}{\Delta\theta}
\label{delta}
\eeq

\noindent
For values of $\delta\ll1$ the particle trajectory is nearly a
straight line (with small perpendicular motions introduced by the
Lorentz forces). In this case the angle averaged spectral energy,
neglecting plasma dispersion, is given by Landau \& Lifshitz (1971,
section 77, p. 215):

\beq
\frac{dW}{d\omega}=\frac{e^2\omega}{2\pi c^3}\int_{\omega/2\gamma^2}^\infty
\frac{\left|{\bf w}_{\omega'}\right|^2}{\omega'^2}
\left(1-\frac{\omega}{\omega'\gamma^2}+\frac{\omega^2}{2\omega'^2\gamma^4}
\right)\,d\omega' 
\label{dW2/dw}
\eeq

\noindent
where ${\bf w}_{\omega'}$ is the Fourier transform of the transverse
acceleration of the particle due to Lorentz forces (${\bf w}\equiv
F_L/\gamma m$).  Fourier transforming the acceleration field along a
particles trajectory ${\bf w}({\bf r}_0+{\bf v}t,t)$ yields

\begin{eqnarray}
{\bf w}_{\omega'}
&=& (2\pi)^{-4}\int e^{i\omega't}\,dt 
\left( e^{-i(\Omega t-{\bf k\cdot r}_0-{\bf k\cdot v}t)} 
{\bf w}_{\Omega,{\bf k}}\, d\Omega d{\bf k} \right)
\nonumber\\
&=&(2\pi)^{-3}\int{\bf w}_{\Omega,{\bf k}}
\delta(\omega'-\Omega+{\bf k\cdot v}) \,e^{i{\bf k\cdot r}_0}
\,d\Omega d{\bf k}.
\end{eqnarray}

\noindent
Squaring the above result and averaging over volume using the results
$\langle|{\bf w}_{\omega'}|^2\rangle =V^{-1}\int|{\bf
w}_{\omega'}|^2d{\bf r}_0$ and $\int e^{i({\bf k-k}_1)\cdot{\bf
r}_0}\,d{\bf r}_0 =(2\pi)^3\delta({\bf k-k}_1)$ yields

\beq
\langle|{\bf w}_{\omega'}|^2\rangle
=(2\pi)^{-3}V^{-1}\int |{\bf w}_{\Omega,{\bf k}}|^2
\delta(\omega'-\Omega+{\bf k\cdot v})\,d\Omega d{\bf k}.
\label{w1}
\eeq

\noindent
In the absence of electric fields\footnote{See \S~2.1 for a discussion
on the implications of this assumption.} the Lorentz acceleration is
given by $(e/\gamma m c){\bf v\times B}$ which, in tensor notation, is
$(e/\gamma m c)\frac{1}{2}e_{\alpha \beta \gamma}(v_\beta B_\gamma
-v_\gamma B_\beta)$.  After simplification this expression results in
\beq |{\bf w}_{\Omega,{\bf k}}|^2 = (ev/\gamma m c)^2 (\delta_{\alpha
\beta} - v^{-2} v_\alpha v_\beta)\, B_{\Omega,{\bf k}}^\alpha
B_{\Omega,{\bf k}}^{*\beta}.
\label{w2}
\eeq

\noindent
where $B_{\Omega,{\bf k}}^\alpha B_{\Omega,{\bf k}}^{*\beta}$ is the
Fourier Transform of the field correlation tensor

\beq
B_{\Omega,{\bf k}}^\alpha B_{\Omega,{\bf k}}^{*\beta}
= T V K_{\alpha \beta}(\Omega,{\bf k})
=TV\!\!\!\int\!\! e^{i(\Omega t-{\bf k\cdot r})} K_{\alpha \beta}({\bf r},t)
\, d{\bf r} dt,
\eeq

\noindent
where T can be interpreted as the period or duration of emission, V is
the volume integrated over and $K_{\alpha \beta}({\bf r},t)$ is the
second order correlation tensor of the magnetic field. Using the
results above and assuming a time independent magnetic field,
equations~(\ref{w1}) and~(\ref{w2}) reduce to

\begin{eqnarray}
&&\langle|{\bf w}_{\omega'}|^2\rangle
=(2\pi V)^{-1}\int|{\bf w}_{\bf k}|^2\delta(\omega'+{\bf k\cdot v})\,
d{\bf k},
\label{w1s}\\
&&|{\bf w}_{\bf k}|^2 
= (ev/\gamma m c)^2 (\delta_{\alpha \beta} - v^{-2} v_\alpha v_\beta)\,
VT K_{\alpha \beta}({\bf k}).
\label{w2s}
\end{eqnarray}

\noindent
The form of $ K_{\alpha \beta}({\bf k})$ used in this paper is taken
from numerical simulations of the Weibel Instability and is given by
\beq K_{\alpha \beta}({\bf k})=C(\delta_{\alpha \beta}-n_\alpha
n_\beta) f_z(k_\|) f_{xy}(k_\perp), \eeq

\noindent
where ${\bf n}$ is the normal to the shock front, C is proportional to
the mean square magnetic field $\left<B^2\right>$, and $f_z(k_\|)$ and
$f_{xy}(k_\perp)$ describe the structure of the magnetic field
parallel and perpendicular to the shocks normal.
\noindent
Inserting the above into equations~(\ref{w1}) and~(\ref{w2}) and
simplifying $ (\delta_{\alpha \beta} - v_\alpha v_\beta/v^2)
(\delta_{\alpha \beta}-n_\alpha n_\beta)=1+(n_\alpha
v_\alpha)^2/v^{2}=1+\cos^2\Theta',$ finally gives the result (for
$v\sim c$; Medvedev 2006)
\begin{eqnarray} 
\langle|{\bf w}_{\omega'}|^2\rangle &=&({\frac{e}{\gamma
m}})^2\frac{CT}{2\pi}\,(1+\cos^2\Theta') \times \nonumber \\ 
&\times &\int\!\! f_z(k_\|)
f_{xy}(k_\perp) \delta(\omega'+{\bf k\cdot v})\,dk_\|d^2 k_\perp.
\label{w2-main}
\end{eqnarray}


\begin{thebibliography}{}

\bibitem{} Akerlof C., et al., 1999, Nature, 398, 400

\bibitem{} Attwood, D., et al. 1993, Appl. Opt., 32, 7022

\bibitem{} Attwood, D., 2000, Soft X-ray and extreme ultraviolet
radiation: principles and applications (Cambridge Univ. Press: New York)

\bibitem{} Bel'kov, S.A., Nikolaev, Y.A., Tsytovich, V.N. 1980,
Sov. Radiophys., 23, 181

\bibitem{} Blandford, R.D. 1972, Astron. Astroph. 20, 135

\bibitem [Blandford \& McKee 1976]{blandfordmckee} Blandford, R. D.,
\& McKee, C. F. 1976, Physics of Fluids, 19, 1130

\bibitem[Chang et al.(2007)]{CSA07}
Chang, P., Spitkovsky, A., Arons, J. 2007, ApJ, submitted, arXiv:0704.3832

\bibitem{} Covino S., et al., 1999, A\&A, 348, L1

\bibitem{} Crider A., et al., 1997, ApJ, 479, L39 

\bibitem{} Ferede, R., et al. 1990, Phys. Scr., 30, 192

\bibitem{} Fleishman, G. D. 2006a, in High-Frequency Waves in
Geospace, ed. J. Labelle \& R. Treumann (Berlin: Springer), 83

\bibitem [(Fleishman 2006)]{fleishman} Fleishman, G. 2006b, ApJ,638, 348

\bibitem [(Frederiksen et al. 2004)] {frederiksenetal} Frederiksen,
J., Hededal, C., Haugb\o lle, T. \& Nordlund, \AA., 2004, ApJ, 608, L13

\bibitem{} Ghisellini G., Celotti A., Lazzati D., 2000, MNRAS, 313, L1

\bibitem{} Ginzgburg, V.L. \& Tsytovich, V.N. 1984, The transition
radiation and transition scattering (Moscow, Russia: Nauka)

\bibitem [(Granot, Piran, and Sari (a) 1999)] {GPS1} Granot, J.,
Piran, T., \& Sari, R. 1999a, ApJ, 513, 679

\bibitem [(Granot, Piran, and Sari (b) 1999)] {GPS2} Granot, J.,
Piran, T., \& Sari, R. 1999, ApJ, 527, 236

\bibitem{} Gunn, J.E. \& Ostriker, J.P. 1971, ApJ, 157, 1395

\bibitem{} Hededal, C., 2005, Ph.D. thesis (astro-ph/0506559)

\bibitem{} Imamura J.~N., Epstein R.~I., 1987, ApJ, 313, 711

\bibitem{} Joshi, C., et al. 1987, IEEE J. Quantum Electron., 23, 1573

\bibitem{} Kaplan, S.A., \& Tsytovich, V.N. 1969, Sov. Phys. Uspekhi, 97, 77
(Usp. Fiz, Nauk, 97, 77) 

\bibitem{} Kincaid, B.M. 1977, J. Appl. Phys., 48, 2684

\bibitem [(Landau \& Lifshitz 1971)] {ll} Landau, L. D., \& Lifshitz,
E.M. 1971, The Classical Theory of Fields (Oxford: Pergamon Press)

\bibitem{} Landau, L.D. \& Pomeranchuk 1953, Sov. Phys. Doklady, 92, 535
(Dokl. Akad. Nauk SSSR, Ser. Fiz. 92, 535

\bibitem{} Lazzati D., et al., 2004, A\&A, 422, 121 

\bibitem [(Medvedev 2000)] {medvedev2000} Medvedev, M. V. 2000, ApJ, 540, 704

\bibitem{} Medvedev. M.V. 2005, astro-ph/0503463

\bibitem [(Medvedev 2006)] {medvedev2006} Medvedev, M. V. 2006, ApJ, 637, 869

\bibitem [(Medvedev \& Loeb 1999)] {medvedev&loeb} Medvedev, M. V., \&
Loeb, A. 1999, ApJ, 526, 697

\bibitem [(Medvedev 2005)] {medvedev2005} Medvedev, M. V., Fiore, M.,
Fonseca, R., Silva, L. O., \& Mori, W. 2005, ApJ, 618, L75

\bibitem [(Medvedev, et al. 2006)] {medvedev2006} Medvedev, M. V.,
Lazzati, D., Morsony, B. C., \& Workman, J. C., 2007, 666, 339

\bibitem{} Meszaros P., Rees M.~J., 1997, ApJ, 476, 232
 
\bibitem{} Migdal, A.B. 1954, Sov Phys. Doklady, 94, 1033
(Dokl. Akad. Nauk SSSR, 94, 1033)

\bibitem[(Nishikawa, et al. 2003)]{nishetal} Nishikawa, K.-I., Hardee,
P., Richardson, G., Preece, R., Sol, H., \& Fishman, G. J. 2003, ApJ,
595, 55

\bibitem{} Panaitescu A., Kumar P., 2001, ApJ, 554, 667

\bibitem{} Panaitescu A., 2005, MNRAS, 363, 1409

\bibitem [(Piran 2004)] {piran2004} Piran, T. 1999, Physics Reports,
Volume 314, Issue 6, 575

\bibitem{} Preece R.~D., Briggs M.~S., Pendleton G.~N., Paciesas
W.~S., Matteson J.~L., Band D.~L., Skelton R.~T., Meegan C.~A., 1996,
ApJ, 473, 310

\bibitem{} Preece R.~D., Briggs M.~S., Mallozzi R.~S., Pendleton
  G.~N., Paciesas W.~S., Band D.~L., 1998, ApJ, 506, L23

\bibitem{} Rees, M.J. 1971a, Nature, 229, 312

\bibitem{} Rees, M.J. 1971b, Nature, 230, 55

\bibitem{} Rossi E., Rees M. J., 2003, MNRAS, 339, 881

\bibitem [(Rybicki \& Lightman 1979)] {r&b} Rybicki, G. B., \&
Lightman, A. P. 1979, Radiative Processes in Astrophysics, (New York:
Wiley)

\bibitem{} Sari R., Piran T., Halpern J.~P., 1999, ApJ, 519, L17

\bibitem [(Silva, et al. 2003)]{silvaetal}
Silva, L. O., Fonseca, R. A., Tonge, J. W., Dawson, J. M.,
Mori, W. B., \& Medvedev, M. V. 2003, ApJ, 596, L121

\bibitem[Spitkovsky(2005)]{Spit05}
Spitkovsky, A. 2005, AIP Conf. Proc., 801, 345;
arXiv:astro-ph/0603211

\bibitem[Spitkovsky(2007)]{Spit07}
Spitkovsky, A. 2007, ApJL, submitted

\bibitem{} Strohmayer T.~E., Fenimore E.~E., Murakami T., Yoshida A.,
1998, ApJ, 500, 873

\bibitem{} Toptygin, I.N. 1985, Cosmic rays in interplanetary 
magnetic fields, (Dordrecht, Netherlands: D. Reidel Publ. Co.)

\bibitem{} Toptygin, I. N., \& Fleishman, G. D. 1987a, Ap\&SS, 132, 213

\bibitem{} Toptygin, I. N. \& Fleishman, G. D. 1987b,
Radiophys. Quantum Electron., 30, 551

\bibitem{} Wang, S., et al. 2002, PRL, 88, 135004-1

\bibitem [(Weibel 1959)] {weibel} Weibel, E. S. 1959, Phys. Rev. Lett., 2, 83

\bibitem{} Wiersma J., Achterberg A., 2004, A\&A, 428, 365
 
\bibitem{} Wijers R.~A.~M.~J., Galama T.~J., 1999, ApJ, 523, 177

\bibitem{} Williams, R.L., et al. 1993, IEEE Trans. Plasma Sci., 21, 156

\bibitem [(Zhang \& M\'esz\'aros 2004)] {z&m} Zhang, B. \&
M\'esz\'aros 2004, International Journal of Modern Physics, 19, 2385

\end{thebibliography}
\end{document}